\documentstyle[aps,prb,psfig,floats]{revtex}
\begin{document}
\draft

\twocolumn[\hsize\textwidth\columnwidth\hsize\csname %
@twocolumnfalse\endcsname

\title{Absence of non-linear Meissner effect in YBa$_2$Cu$_3$O$_{6.95}$ }
\author{A. Carrington$^{a,b}$ , R. W. Giannetta$^a$, J.T. Kim$^a$, and J. Giapintzakis$^{a,*}$}
\address{$^a$ Department of Physics, University of Illinois at Urbana-Champaign, \\ 1110 West Green St, Urbana, 61801 Illinois, USA.} 
\address{$^b$ Department of Physics and Astronomy, University of Leeds, Leeds LS2 9JT, England.}
\date{\today}
\maketitle

\begin{abstract}
We present measurements the field and temperature dependence of the penetration depth ($\lambda$) in high purity, untwinned single crystals of YBa$_2$Cu$_3$O$_{6.95}$ in all three crystallographic directions.  The temperature dependence of $\lambda$ is linear down to low temperatures, showing that our crystals are extremely clean.  Both the magnitude and temperature dependence of the field dependent correction to $\lambda$ however, are considerably different from that predicted from the theory of the non-linear Meissner effect for a $d$-wave superconductor (Yip-Sauls theory).  Our results suggest that the Yip-Sauls effect is either absent or is unobservably small in the Meissner state of YBa$_2$Cu$_3$O$_{6.95}$.
\end{abstract}
\pacs{PACS numbers: 74.72.Bk,74.25.Nf }
]


The identification of a strong linear term in the temperature dependence of the magnetic penetration depth ($\lambda$) of YBa$_2$Cu$_3$O$_{6.95}$ was a key development in identifying the $d$-wave symmetry of the order parameter in high temperature superconductors. \cite{hardy93}
The linear increase of $\lambda$ with increasing temperature in a $d$-wave superconductor reflects a decrease in superfluid density caused by thermal excitations near the nodes in the order parameter, and is insensitive to the direction of the superflow.  Yip and Sauls \cite{yip92,xu95} have proposed that a much more sensitive test of the order parameter symmetry can be made by measuring the field dependence of $\lambda$ in the Meissner state.  They argued that at sufficiently low temperature, a pure $d$-wave superconductor should have a weak, linear field dependence, the magnitude of which depends on the relative orientations of the applied field and the nodes in the order parameter.

Although the linear temperature dependence of $\lambda$ is now well established in many materials the field dependence has proved much more difficult to isolate.  There is to date no conclusive evidence as to the presence or absence of this effect in the Meissner state in any material.  In this paper we present measurements of both the field and temperature dependence of $\lambda$ below $H_{c1}$ in high purity, untwinned single crystals of YBa$_2$Cu$_3$O$_{6.95}$, in all three crystallographic directions ($H \| a, b, c$).  We find that in some crystals there is a relatively strong field dependence to $\lambda$ which might naively be attributed to the Yip-Sauls effect. However, a closer examination reveals that it is almost certainly of extrinsic origin. Our cleanest crystal shows a field dependence of $\lambda$ which is up to a factor 7 lower than that expected from the Yip-Sauls theory and is not suppressed by increasing temperature.  This suggests that the Yip-Sauls effect is either absent or unobservably small in the Meissner state of YBa$_2$Cu$_3$O$_{6.95}$.

In the Yip-Sauls theory the field dependence of $\lambda$ at low temperature is a direct consequence of the presence of the nodes in the order parameter.  In the presence of a finite superflow ($\vec{v}_s$) the quasiparticle energy levels are Doppler shifted by an amount $\delta \epsilon \propto \vec{v}_s \cdot \vec{v}_f$, where $\vec{v}_f$ is the Fermi velocity.  In a $d$-wave superconductor, levels near a node with $\vec{v}_f$ in the opposite direction to $\vec{v}_s$ will be shifted below the Fermi level and there will be a transfer of quasiparticles from comoving to countermoving states, thus resulting in a reduction of the effective superfluid density.  In an $s$-wave superconductor the effect is absent at low temperature because the finite gap prevents the shifted levels from being occupied.  Yip-Sauls calculate the field dependence in a $d$-wave superconductor to be
\begin{equation}
\frac{\Delta\lambda(H)}{\lambda (0) } =\alpha \frac{H}{H_0}
\end{equation}
where $H_0$ is of order the thermodynamic critical field (Xu {\it et al.} \cite{xu95} estimated it to be 2.5T for YBa$_2$Cu$_3$O$_{6.95}$), and $\alpha$ depends on the relative orientation of the nodes and the field.  For field directed along an antinode, we calculate from this formula that the slope $d\lambda_a(H)/dH = 4.7 \times 10^{-2}$ \AA/Oe for YBa$_2$Cu$_3$O$_{6.95}$, \cite{caldlh} or 4.7 \AA\ in a field of 100\ Oe.  The Yip-Sauls effect is very sensitive to temperature and sample purity.  At finite temperature $\Delta\lambda$ is only linear in $H$ above a crossover field $H^* \simeq k_bTH_0/2\Delta_0$ which exceeds $H_{c1}$ for $T\gtrsim3$ K (here $\Delta_0$ is the maximum energy gap $\sim 1.9 T_c$.\cite{xu95}  Below $H^*$ $\Delta\lambda\sim H^2$ and $d\lambda/dH$ is much reduced.  This high sensitivity on temperature can be used to distinguish the Yip-Sauls effect from other possible (non-intrinsic) contributions.

Our measurements of $\lambda$ were performed by measuring sample induced changes in the inductance of a copper coil (containing the sample).  The coil formed part of a tunnel diode driven LC oscillator operating at 13 MHz.  Changes in the resonant frequency of this circuit are directly proportional (to within a calibration constant) to changes in the effective volume of the inductor.  By careful design of this measurement circuit (following the prescription given by Van Degrift \cite{degrift75}), we have achieved a RMS noise level in the oscillator of 1 part in $10^9 /\surd$Hz.  For a typical crystal with 1mm$^2$ surface area, this translates to a noise level in the measurement of $\lambda$ of $\sim$ 0.1\AA.  This resolution is about 100 times better than previous measurements. \cite{maeda95}

A small superconducting solenoid coaxial with the probe coil provides a variable, static, dc field, up to $\sim$1000 Oe collinear with the RF probe field. The complete experiment was placed inside a mumetal shielded Dewar.  The remnant field inside this was measured, using a fluxgate magnetometer, to be $\lesssim$ 2 mOe.  The RF probe field to which the sample is subjected is estimated to be also  $\lesssim$ 2 mOe.  For a single crystal, with the $H$ parallel to the plane, the total flux in the sample is of order one flux quantum ($\Phi_0$).  These very low field values help keep to an absolute minimum the number of vortices trapped in the sample as it is cooled through $T_c$.  

The sample is located on a sapphire rod, the temperature of which is varied independent of the rest of the apparatus.   The sample rod may be drawn out of the measurement coil at low temperature so that the background field dependence of the apparatus can be measured {\it in-situ}.  We found that this background was highly reproducible at low temperature, but could change markedly upon thermal cycling, making an {\it in-situ} measurement vital.  There was a further background originating from the sapphire sample holder, which was temperature dependent, but highly reproducible between runs.  This was therefore measured in a separate run with the sample removed.  These backgrounds typically  give a frequency shift corresponds to around a 3 \AA\ equivalent change in $\lambda$ for a mm sized single crystal in 100\ Oe.

The effective volume of the sense coil was calculated by measuring the shift in frequency when a small sphere of Aluminium was inserted.  In the $B\| a,b$ geometry the measured frequency shifts are then easily related to changes in penetration depth simply by measuring the area of the faces of the sample.  This neglects the contribution of the currents flowing along the $c$-axis, although for the aspect ratio of the crystals measured here, this results in less than a 1\% error in $\Delta\lambda(T=30K)$. \cite{bonn96}  For the $B\| c$ geometry a calibration factor is rather more difficult to calculate and so it was estimated by comparing the temperature dependence of the frequency shifts in this orientation to that for $B\| a,b$.  This factor was then used to relate the field dependent shifts to $\Delta\lambda(H)$. 

The single crystals measured were grown in Yttria stabilized Zirconia crucibles as described in Ref.\ \onlinecite{stupp92}.  They were annealed so as to give optimal doping.  The crystal used for the bulk of the measurements described here had dimensions $a\times b \times c$ = $0.80 \times 0.51 \times 0.011$ mm$^3$, was naturally untwinned  and had a $T_c$ of 91.4K.  The crystal contained a single twin boundary located on one corner; this region of material with flipped $a,b$ axes was only $\sim$ 5\% of the total crystal volume.  X-ray diffraction was used to identify the crystal axes as well as to confirm the (essentially) untwinned nature of the crystal. 

\begin{figure}
\centerline{\psfig{figure=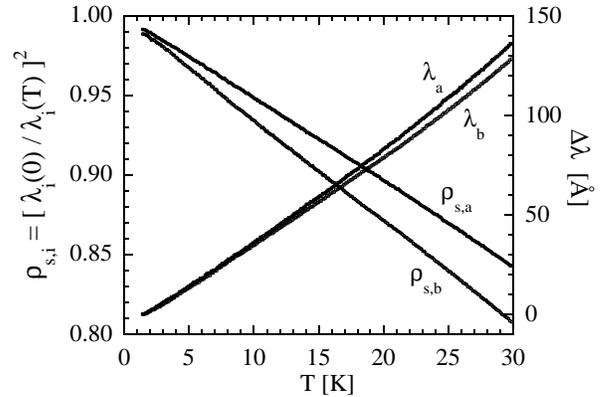,width=8cm}}
\caption{Temperature dependence of $\lambda_a$ and  $\lambda_b$ for our untwinned single crystal of YBa$_2$Cu$_3$O$_{6.95}$, in zero dc field. The superfluid density [$\rho_{s,i}=(\lambda_i(0)/\lambda_i(T))^2$] is also shown.}
\label{tdepfig}
\end{figure}

The temperature dependence of the penetration depth and the superfluid density [$\rho_{s,i}=(\lambda_i(0)/\lambda_i(T))^2$] for the field applied along the $a$ and $b$ axes is shown in Fig.\ \ref{tdepfig}. 
$\lambda(0)$ values were taken from  Bazov {\it et al.} \cite{bazov95,deltalambdanote} ($\lambda_a(0)=1600$\AA\  and $\lambda_b(0)=$1200 \AA).  It can been seen that $\rho_s$ is essentially linear down to the lowest temperature measured (1.4 K).  A fit to the dirty $d$-wave interpolation formula of Hirschfeld and Goldenfeld \cite{hirschfeld93} [$\lambda(T) = a T^2/(T+T^*)$] yields values of $T^* \lesssim 1$K for both directions.  We note that the measured anisotropy in $d\lambda/dT$ at low temperatures is only $\sim $ 3\% ($d\lambda_a/dT$ = 4.34 \AA/K and $d\lambda_b/dT$ = 4.23 \AA/K) which is somewhat less than that reported by Bonn {\it et al.}\cite{bonn96}  This may reflect a lower anisotropy in the zero temperature values of $\lambda$ in our crystals.  We note that our values for $d\rho_s/dT$ are within 5\% of those in Ref.\ \onlinecite{bonn96}.  The values of $T^*$ for this crystal are among the lowest reported to date, showing that our crystal is extremely clean.  This property is extremely important as the Yip-Sauls theory predicts that even small amounts of impurities depress the field dependence of $\lambda$.

A major problem in measuring the field dependence of $\lambda$, by the method described above, is isolating changes in the London penetration depth ($\lambda_L$) from other effects which also increase the effective penetration depth  $\lambda_E$.  The most obvious extrinsic contribution is caused by the motion of vortices.  In the mixed state the additional field penetration caused by vortex motion is given by Campbell penetration depth $\lambda_{p} = B \phi_0/\mu_0 \kappa_p$  ($\kappa_p$ is the pinning force, and $\phi_0$ is the flux quantum).  In the limit that $\lambda_p$ is small compared to $\lambda_L$, the contributions add in quadrature, \cite{coffey92} $\lambda_E^2(H)=\lambda_L^2(H)+\lambda_p^2$ and so 
\begin{equation}
\Delta\lambda_E \simeq \Delta\lambda_L+\frac{B\phi_0}{2\mu_0\lambda_L\kappa_p}, \quad\quad \lambda_p\ll\lambda_L
\end{equation}
Using values of $\kappa_p$ taken from Ref.\ \onlinecite{wu90} we find that at low temperature the vortex contribution would be $\sim$ 25\AA\ in 100\ Oe (for $H\|ab$) if $H_{c1}$ was zero and flux was free to enter the sample. For $H$ less than the actual $H_{c1}$ the sample is in the Meissner state and should, in principle, be free of vortices.  However, in practice vortices may enter well before $H_{c1}$ at sharp corners and surface imperfections and so in these regions $\lambda_E>\lambda_L$.  In the measurement configuration $H\|c$ this problem is likely to be intensified as the field lines are highly distorted due to the flat plate geometry of the sample.  Another contribution may come from weak links at the surface.  An applied field may dramatically suppress the critical current of any weak link thus causing an increase in $\lambda_E$.  Finally in the configuration $H\|ab$ part of the supercurrent flows along the $c$-axis, and  $\lambda_E \propto \lambda_{ab}+t/l \lambda_c$ (where $t$ and $l$ are the dimensions of the sample along the $c$ and $a,b$ axes respectively).  As already mentioned, because of the relative weak anisotropy of YBa$_2$Cu$_3$O$_{6.95}$ this introduces only a small correction to the measured temperature dependence of $\lambda$ for thin samples.  However, the contribution of $\lambda_c$ to the field dependence of  $\lambda$  may not be negligible.   From the above it is clear that the measured field induced changes in $\lambda$ can only set an upper limit for the intrinsic field dependence of $\lambda_L$, in all field configurations.

\begin{figure}
\centerline{\psfig{figure=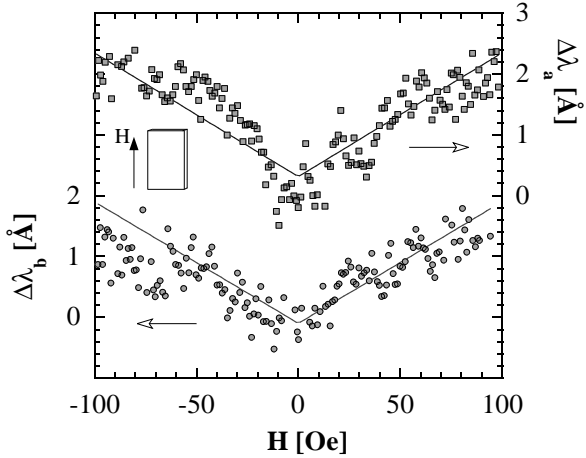,width=8cm}}
\caption{Field dependence of $\lambda$ in the $a$ and $b$ directions at 1.4 K.}
\label{abfieldfig}
\end{figure}

The field dependence of $\lambda$ for our YBa$_2$Cu$_3$O$_{6.95}$ crystal in the $a$ and $b$ directions at our lowest temperature (1.4K) is shown in Fig.\ \ref{abfieldfig}.  Although $\Delta\lambda(H)$ is approximately linear with field its magnitude is much smaller than the Yip-Sauls prediction.  The slope of the straight line fitting through that data is $20\pm5 \times 10^{-3}$ \AA/Oe (i.e., $\sim$ 2\AA\ in 100 Oe), and within the error is the same in both directions.  This is more than a factor 2 times lower than the Yip-Sauls prediction.  No hysteresis was seen within the scatter of the data. 

\begin{figure}
\centerline{\psfig{figure=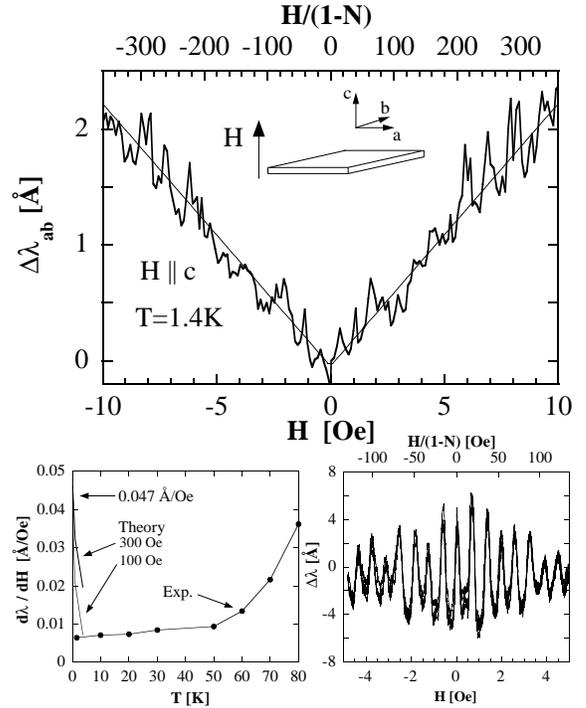,width=8cm}}
\caption{{\it Top}: Field dependence of $\lambda_{ab}$ with the field applied along $c$ at 1.4K.  {\it Bottom left}: The evolution of $d\lambda_{ab}/dH$ with temperature.   Theory curves are the average slope $\Delta\lambda(H)/H$ with $H=100$ Oe and 300 Oe from Ref.\ \protect\onlinecite{xu95}. {\it Bottom Right}: Fraunhofer like oscillations in $\lambda(H)$ in another sample.  The data are highly reproducible (5 field sweeps are shown).}
\label{cfieldfig}
\end{figure}

Data for the changes in the average in-plane penetration depth ($\lambda_{ab}$) in the third field configuration ($H\|c$) are shown in the upper panel of Fig.\ \ref{cfieldfig}.  $\Delta \lambda_{ab}(H)$ is linear up to the maximum field measured [a decrease in slope was seen above $\sim$ 20 Oe (not shown)].  Again no hysteresis was observed at these low field values. For this flat sample, in this geometry, the field at the surface of the sample is clearly very different from the applied field.  We have estimated the field at the edge of the sample by multiplying the applied field by a demagnetizing factor [1/(1-N)], derived from an inscribed ellipsoid approximation. This {\lq}demagnetizing{\rq} factor was within 20\% of that estimated by comparing the known volume of the sample to the measured apparent shielding volume (as measured in a zero field cooled dc magnetization measurement).  The actual field on the edge of the sample will not be constant over the whole width, however this {\lq}demagnetizing{\rq} factor is a rough estimate of the average field.  Using 1/(1-N)=36, we estimate the slope $d\lambda_{ab}/dH = 6.3 \times 10^{-3}$ \AA/Oe, which is $\sim$ 3 times smaller than for the $H\|a,b$ measurements and more than 7 times smaller than the Yip-Sauls prediction.  We note that with this demagnetizing factor that the change in slope at H$\sim$ 20 Oe corresponds roughly to $H_{c1}$ in this direction.

The lower left panel of Fig.\ \ref{cfieldfig} shows the evolution  of the slope, $d\lambda_{ab}/dH$ as temperature is increased.  There is almost no change in $d\lambda_{ab}/dH$ up to 50K, in sharp contrast to the predictions of the Yip-Sauls theory.  As $T_c$ is approached $d\lambda_{ab}/dH$ increases substantially.  There was no indication of the predicted change from $\Delta\lambda (H)\propto H$ to $\Delta\lambda (H)\propto H^2$ as temperature was increased.  We conclude from these observations that the small field dependence that we see does not originate from the Yip-Sauls effect.  If we interpret $\Delta\lambda(H)$ as originating from vortex motion, the above results would imply that at low temperature $\kappa_p$ is almost constant in our samples.

A close examination of $\lambda_{ab}(H)$ in Fig.\ \ref{cfieldfig} shows that the apparent noise on top of the linear dependence is reproducible with a pattern which is approximately symmetric about $H=0$.  In fact these modulations in $\lambda_{ab}$ are quite reproducible and become more pronounced as temperature is increased.  In another sample we observed much bigger modulations which resembled closely the type of Fraunhofer pattern typical of the field dependence of the critical current in weak links (see the lower right panel in Fig.\ \ref{cfieldfig}).  We conclude that these effects indicate the presence of weak links on the edge of the crystal, the critical current of which is modulated by the static dc field.  We note that this structure is only observable because of the extremely low modulation fields used in our experiment.

\begin{figure}
\centerline{\psfig{figure=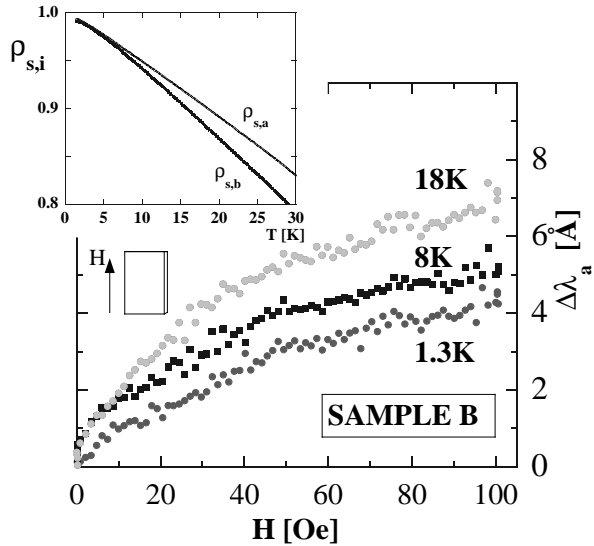,width=8cm}}
\caption{Field dependence of $\lambda_a$ for a second sample, at several fixed temperatures (indicated on figure). $H$ was $\|b$.  The inset shows the temperature dependence of $\rho_{s,a}$ and $\rho_{s,b}$ for the same sample.}
\label{samplebfig}
\end{figure} 

Finally in Fig.\ \ref{samplebfig} we show data for a second sample of optimally doped, untwinned YBa$_2$Cu$_3$O$_{6.95}$ ($T_c$=93.6 K).   $\rho_s(T)$ (inset Fig.\ \ref{samplebfig}) for this sample is linear down to only $\sim$4K, showing that it is dirtier than the first sample (cf., Fig.\ \ref{tdepfig}).  $\Delta \lambda (H)$ however is larger by around a factor 2 and is linear up to a field of approximately 50 Oe.  Again no hysteresis was observed over the field range in the figure ($H<$100 Oe).  This linear slope increases slightly with increasing temperature in sharp contrast to the predictions of the Yip-Sauls theory.  From this fact we conclude that this linear contribution is again of extrinsic origin.  The larger magnitude of $\Delta\lambda(H)$ probably reflects more flux leaking into this sample.

The weak temperature dependence of  $d\lambda_{ab}/dH$ shown here is consistent with the results of Maeda {\it et al.} \cite{maeda95}  for their samples of YBa$_2$Cu$_3$O$_{6.95}$ and Bi$_2$Sr$_2$CaCu$_2$O$_8$.  In their work the lowest temperature at which experiments were performed was 10K, and little change was found in $d \lambda/dH$ as temperature was increased.  However these authors report much bigger changes in $\lambda(H)$ than are found here or are indeed predicted by the Yip-Sauls theory. The large size of these effects and the fact that they are not damped out by increasing temperature leads us to conclude that they do not originate from the Yip-Sauls effect, and probably arise from vortex motion.

In summary, we have presented measurements of the field dependence of $\lambda$ with a resolution around a factor 100 better than previous experiments, on very high purity single crystals of YBa$_2$Cu$_3$O$_{6.95}$, at temperatures down to $T/T_c$=0.015.  In our cleanest sample we observe a very weak field dependence in all three crystallographic directions.  The overall magnitude of this change in $\lambda$ is at least two times smaller ( or 7 times smaller in the $H\|c$ configuration) than our estimates of what we would expect to see if the Yip-Sauls theory was valid.  A second sample shows a larger effect even though its purity level is lower.  In all cases however the observed field dependence of $\lambda$ is not diminished by increasing temperature and so we conclude that it does not arise from the Yip-Sauls effect.  The sample dependent $\Delta \lambda (H)$ we observe most likely originates from vortex motion.

The large body of evidence collected thus far shows fairly convincingly that YBa$_2$Cu$_3$O$_{6.95}$ is a $d$-wave superconductor and so other possible reasons must be considered for the apparent absence of the Yip-Sauls effect below $H_{c1}$.  One possibility could be the effect of non-local electrodynamics.  It is usually assumed that as the in-plane penetration depth ($\lambda_0 \sim$1600\AA) is much greater than the in-plane coherence length ($\xi_0 \sim 16$\AA) then local electrodynamics are sufficient.  Kosztin {\it et al.} \cite{kosztin97} however have pointed out that this argument does not hold for a $d$-wave superconductor, as $\xi$ is angle dependent. At the nodes the energy gap $\Delta(\theta)\propto\xi(\theta)^{-1}$ goes to zero and $\xi(\theta)$ diverges violating the local condition.  As the region near to the nodes is exactly where the low temperature excitations are formed, this observation can be expected to have serious consequences for both the temperature and field dependence of $\lambda$.   Li {\it et al.} \cite{li98} have recently made calculations of the $\Delta \lambda(H)$ which include both the non-local and non-linear effects.  They conclude that the non-local effects indeed drastically reduce $\Delta \lambda(H)$ at fields $\lesssim H_{c1}$, rendering the effect essentially unobservable in the Meissner state.  Although these calculations strictly only apply in the $H\| c$ configuration (when the direction of field penetration is in the same plane as the supercurrents), the relatively low anisotropy of YBa$_2$Cu$_3$O$_{6.95}$ means that a similar reduction may also occur in the $H\|$ plane configuration.\cite{li98} 

We thank P.J.\ Hirschfeld and M.R.\ Li for discussing their calculations prior to publication.  We also acknowledge useful conversations with D.M.\ Ginsberg, A.\ Goldman,  I.\ Kosztin, A.J.\ Leggett,and  J.A.\ Sauls.  This work was supported by the National Science Foundation (DMR 91-20000) through the Science and Technology Center for Superconductivity, and by DOE grant DEFG02-91-ER45439.


\begin{references}
\bibitem[*]{} Present address: Laboratory of Superconductivity, FO.R.T.H. - I.E.S.L., 711 10 Heraklion, Greece.

\bibitem{hardy93} W.N. Hardy, D.A. Bonn, D.C. Morgan, R.X. Liang and K. Zhang, Phys. Rev. Lett. {\bf 70}, 3999 (1993).

\bibitem{yip92} S.K. Yip and J.A. Sauls, Phys. Rev. Lett. {\bf 69}, 2264 (1992).

\bibitem{xu95} D. Xu, S.K. Yip and J.A. Sauls, Phys. Rev. B {\bf 51}, 16233 (1995).

\bibitem{caldlh} Yip {\it et al.}\protect\cite{yip92,xu95} define $\lambda=-[d\ln B/dx]^{-1}$.  However, in our experiment the changes in the effective $\lambda$ are related to $\int B dV$, which reduces their $\Delta\lambda(H)$ values  by 2.

\bibitem{degrift75} C.T. Van Degrift, Rev. Sci. Inst. {\bf 46}, 599 (1975).

\bibitem{maeda95} A. Maeda {\it et al.}, Phys. Rev. Lett. {\bf 74}, 1202 (1995); Physica C {\bf 263}, 438 (1996);  J. Phys. Soc. Jap. {\bf 65}, 3638 (1996).

\bibitem{bonn96} D.A. Bonn {\it et al.} Czech. J. Phys. {\bf 46}, 3195 (1996).

\bibitem{stupp92} S.E. Stupp and D.M. Ginsberg, in {\it Physical Properties of High Temperature Superconductors III}, edited by D. M. Ginsberg (World Scientific, Singapore, 1992).

\bibitem{bazov95} D.N. Basov {\it et al.}, Phys. Rev. Lett. {\bf 74}, 598 (1995).

\bibitem{deltalambdanote} $\Delta\lambda(T) = \Delta\lambda(T)+\lambda(T=1.45$K).  We estimated $\lambda(T=1.45$K) by linearly extrapolating our  $\Delta\lambda(T)$ data to $T=0$.

\bibitem{hirschfeld93} P.J. Hirschfeld and N. Goldenfeld, Phys. Rev. B {\bf 48}, 4219 (1993).

\bibitem{coffey92} M.W. Coffey and J.R. Clem, Phys. Rev. B {\bf 45}, 9872 (1992).

\bibitem{wu90} D.H. Wu and S. Sridhar, Phys. Rev. Lett. {\bf 65}, 2074 (1990).

\bibitem{kosztin97} I. Kosztin and A.J. Leggett, Phys. Rev. Lett. {\bf 79}, 135 (1997).

\bibitem{li98} M.-R. Li, P.J. Hirschfeld, and P. W\"olfle, Phys. Rev. Lett. {\bf 81}, xxxx (1998). 

\end{references}
\end{document}